\documentclass{article}

\usepackage{amssymb} 
\usepackage[colorlinks,linkcolor=green,anchorcolor=red,citecolor=blue]{hyperref}

\begin{document}

\title{Response to Comments on ``PCA Based Hurst Exponent Estimator for
fBm Signals Under Disturbances''}

\author{Li~Li}

\maketitle

\begin{abstract}
In this response, we try to give a repair to our previous proof
given in Appendix of \cite{LiHuChenZhang2009} by using orthogonal
projection. Moreover, we answer the question raised in
\cite{Zanten2010}: If a centered Gaussian process $G_t$ admits two
series expansions on different Riesz bases, we may possibly study
the asymptotic behavior of one eigenvalue sequence from the
knowledge on the asymptotic behaviors of another.
\end{abstract}

\section{The Backgrounds}
\label{sec:1}

Many thanks to the note of Prof. Zanten \cite{Zanten2010}, a flaw
was found to lie in Appendix of Li, et al. \cite{LiHuChenZhang2009},
which tries to give another proof for the asympotitics of the
eigenvalues for Karhunen-Lo\`{e}ve expansion of fBm process.
Fortunately, all the theorems in the mainbody of
\cite{LiHuChenZhang2009} still holds, due to the nice proof of
\cite{Bronski2003a}-\cite{Bronski2003b}. In the rest of this
response, we will try to fix our uncompleted proof in
\cite{LiHuChenZhang2009} and answer a related question raised in
\cite{Zanten2010}.

\indent

Let us briefly recall some backgrounds of our discussions. Suppose
$B = \{ B_t, t \geq 0 \}$ is a standard fBm process within a finite
time interval $[0, 1]$ (can be scaled to $[0, T]$, but it does not
matter our proof) and with Hurst exponent $H$ ($0<H<1$).

The autocorrelation function of $B_t$ can be written as
\cite{MandelbrotvanNess1968}
\begin{equation}
\label{equ:1} R_b(s,t) = E \left[ B_{s} B_{t}\right] = \frac{1}{2}
\left( {s}^{2H} + {t}^{2H} - \left| s-t \right|^{2H} \right)
\end{equation}

According to Mercer's theorem
\cite{Mercer1909}-\cite{HislopSigal1995}, we have
\begin{equation}
\label{equ:2} R_b(s,t) = \sum_{n=1}^{\infty} \lambda_n \phi_n(s)
\phi_n(t)
\end{equation}
%
\begin{equation}
\label{equ:3} \int_0^1 R_b(s,t) \phi_n(t) dt = \lambda_n \phi_n(s)
\end{equation}
\noindent where $\{ \phi_n(t) \}_{n=1}^\infty$ is a set of
orthonormal functions in the interval $[0,1]$, where $\lambda_n$ are
the corresponding eigenvalues of the $n$th orthonormal functions.

As shown in \cite{Karhunen1947}-
\cite{LiHuChenZhang2009}, Eq.(\ref{equ:3}) is the continuous
Karhunen-Lo\`{e}ve (K-L) expansion for fBm process and $\lambda_n$
is the associated eigenvalues. Our main problem here is to discuss
the asympotics of $\lambda_n$.

\section{Prof. Bronski's Proof}
\label{sec:1}

In \cite{Bronski2003a}-\cite{Bronski2003b}, Prof. Bronski had showed
that $\lambda_n \sim n^{-2H-1}$ as follows.

Clearly, from Eq.(\ref{equ:2})-(\ref{equ:3}), we get a integral
kernel $\left[ T_\kappa \phi \right] (x) = \int_0^1 R_b (x, y)
\phi(y) dy$ on $L_2([0, 1] \times [0, 1])$. Moreover, this operator
$T_\kappa$ is a non-negative symmetric, Hilbert-Schmidt and compact.

We can then prove the rigorous estimates of the eigenvalues by
considering the Nystr\"{o}m approximation of this kernel \cite{Nystrom1930}-
\cite{PorterStirling1990} on a special orthonormal basis $\phi_n(x)
= \{\sqrt{2} \sin((n+\frac{1}{2}) \pi x)_{n=0}^{\infty} \}$.
Particularly in \cite{Bronski2003a}-\cite{Bronski2003b}, the
operator $T_\kappa$ is approximated by an operator $A$ from the
sequence space $l_2$ to $l_2$, which has matrix elements
\begin{equation}
\label{equ:4} A_{n,m} = \langle \phi_n(x) A \phi_m(y) \rangle = 2
\int_0^1 \int_0^1 R_b(x,y) \sin((n+\frac{1}{2}) \pi x)
\sin((m+\frac{1}{2}) \pi x) dx dy
\end{equation}

We can also consider $x^T A y$ with the kernel matrix $A_{n,m}$ as a
$n$-degenerate approximation of the Mercer kernel function $R_b (x,
y)$.

By examining the leading order diagonal piece $D$ and the higher
order off-diagonal piece $OD$ of $A$ ($A = D + OD$), Bronski proved
that $OD_{n,m}$ has higher order and can be neglected with respect
to $D_{n,m}$. Thus, the eigenvalues of $A$ is mainly determined by
$D$.

In \cite{Bronski2003a}-\cite{Bronski2003b}, Bronski further proved
that
\begin{equation}
\label{equ:5} \frac{\sin( \pi H) \Gamma(2H+1)}{n^{2H+1}} +
\epsilon_{left} \le \lambda_n (A) \le \frac{\sin( \pi H)
\Gamma(2H+1)}{n^{2H+1}} + \epsilon_{right}
\end{equation}
\noindent where $\epsilon_{left}$ and $\epsilon_{right}$ are
neglectable items.

Thus, we reach the conclusion we desired
\begin{equation}
\label{equ:6} \lambda_n (T_\kappa) \approx \lambda_n (A) \sim
n^{-2H-1}
\end{equation}

\section{Another Proof that Actually Detours}
\label{sec:3}

\subsection{The Proof in Appendix of \cite{LiHuChenZhang2009}}
\label{sec:3.1}

In \cite{LiHuChenZhang2009}, we consider the series expansion of the
fBm process on a set of orthonormal basis functions $\phi_n(t)$
\begin{equation}
\label{equ:7} B_t = \sum_{n=1}^{\infty} c_n \phi_n(t)
\end{equation}
\noindent where $c_n$ is the corresponding coefficient satisfying
\begin{equation}
\label{equ:8} E \{ c_n c_m \} = \lambda_n \delta[n-m]
\end{equation}

If we can obtain the representation of $c_n$, we can directly get
$\lambda_n$ via Eq.(\ref{equ:5}).

However, in \cite{LiHuChenZhang2009} (or equivalently
\url{http://arxiv.org/abs/0805.3002v1}), we instead study another
series expansion of the fBm process on a set of special basis
functions $\psi_n(t)$ proposed in \cite{DzhaparidzeaZanten2004}-
\cite{DzhaparidzeaZanten2005b}
\begin{equation}
\label{equ:9} B_t = \sum_{n=1}^{\infty} b_n \psi_n(t)
\end{equation}
\noindent where $\{ \psi_n(t) \}_{n=1}^\infty$ is a set of linearly
independent but not orthogonal basis functions. Thus, the expansion
coefficients $b_n$ is not equivalent to the eigenvalues of the
Karhunen-Lo\`{e}ve expansion.

The Appendix of \cite{LiHuChenZhang2009} proves the $E \{ b_n \}
\sim n^{-2H-1}$. But as shown in \cite{Zanten2010}, it is just a
intermediate result for our final goal.

\subsection{A Remedy}
\label{sec:3.2}

Because the problem lies in the orthogonality, we will give a remedy
for our proof by orthogonal projection.

More precisely, we will project these functions $\{ \psi_n(t)
\}_{n=1}^\infty$ to a set of orthonormal basis functions $\{
\phi_n(t) \}_{n=1}^\infty$ as
\begin{equation}
\label{equ:10} \psi_n(t) = \sum_{k=1}^{\infty} \mu_{n,k} \phi_k(t)
\end{equation}
\noindent where $\mu_{n,k}= \textrm{proj}_{\phi_n(t)} (\psi_k(t)) =
\left(\int_0^1 \phi_n(t)\psi_k(t) dt \right)$.

Based on Eq.(\ref{equ:7}), (\ref{equ:9})-(\ref{equ:10}), we have
\begin{equation}
\label{equ:11} B_t = \sum_{n=1}^{\infty} c_n \phi_n(x) =
\sum_{n=1}^{\infty} b_n \psi_n(x) = \sum_{n=1}^{\infty}
\left(\sum_{k=1}^{\infty} \mu_{n,k} b_k \right) \phi_n(x)
\end{equation}

Thus, we can study the eigenvalue asymptotics of $c_n$ from
\begin{equation}
\label{equ:12} c_n = \sum_{k=1}^{\infty} \mu_{n,k} b_k
\end{equation}

This method is similar to what had been applied in
\cite{AnguloRuiz-Medina1998}-\cite{Gnaneshwar2007}. We will discuss
when such projection is valid at the end of this response.

The success of Prof. Bronski \cite{Bronski2003a}-\cite{Bronski2003b}
inspired us to choose the orthonormal basis functions $\phi_n(t) =
\{\sqrt{2} \sin((n+\frac{1}{2}) \pi t)_{n=0}^{\infty} \}$. Because
the Karhunen-Lo\`{e}ve expansion for brownian motion $H =
\frac{1}{2}$ is well known \cite{Yaglom1987},
\cite{DzhaparidzeaZanten2004}, we will focus on the cases that $H
\ne \frac{1}{2}$.

As pointed out in \cite{LiHuChenZhang2009}-
\cite{DzhaparidzeaZanten2005b}, we can expand a standard fBm process
$B_t$ as
\begin{equation}
\label{equ:13} B_t = \sum_{n=1}^{\infty} \left( z_n \frac{\sin(x_n
t)}{x_n} + w_n \frac{1 - \cos(y_n t)}{y_n} \right)
\end{equation}
\noindent where $x_n$ are the positive zeros of the Bessel function
$J_{-H}$ of the first kind, $y_n$ are the positive zeros of the
Bessel function $J_{1-H}$ of the first kind. As shown in
\cite{LiHuChenZhang2009}, we have
\begin{equation}
\label{equ:14} x_n = n \pi + h_1 + O(n^{-1}), \indent y_n = n \pi +
h_2 + O(n^{-1})
\end{equation}
\noindent where $h_1$ and $h_2$ are constants.

$z_n$ and $w_n$ are independent sequences of independent, centered
Gaussian random variables on a common probability space, with
\begin{equation}
\label{equ:15} E \left[ z_n \right] = E \left[ w_n \right] = 0
\end{equation}
%
\begin{equation}
\label{equ:16} E \left[ z_n^2 \right] = \frac{2 c_H^2}{x_n^{2H}
J_{1-H}^2(x_n)}, \indent E \left[ w_n^2 \right] = \frac{2
c_H^2}{y_n^{2H} J_{-H}^2(y_n)}
\end{equation}
\noindent where $c_H^2 = \frac{\Gamma(1+2H) \sin(\pi H)}{\pi}$.

Let us first examine the projection of $\{ \frac{\sin(x_n t)}{x_n}
\}_{n=1}^\infty$. Given $n$, $k \in \mathbb{N}$, we can obtain the
projection coefficients $\hat{\mu}_{n,k}$ as
\begin{eqnarray}
\label{equ:17} & & \hat{\mu}_{n,k} \nonumber \\
& = & \int_0^1 \frac{\sin(x_k t)}{x_k} \sqrt{2} \sin([n - \frac{1}{2}] \pi t) dt \nonumber \\
& = & \frac{\sqrt{2}}{2 x_k } \left[ \frac{\sin(x_k - [n -
\frac{1}{2}] \pi)}{x_k - [n - \frac{1}{2}] \pi}
- \frac{\sin(x_k + [n - \frac{1}{2}] \pi)}{ x_k + [n - \frac{1}{2}] \pi} \right]\nonumber \\
& = & \frac{\sqrt{2} d_1}{x_k } \left( \frac{\left(x_k + [n -
\frac{1}{2}] \pi \right) + \left( x_k - [n - \frac{1}{2}] \pi
\right)}{x_k^2 - [n - \frac{1}{2}]^2 \pi^2} + O(n^{-2}) \right) \nonumber \\
& = & \frac{2 \sqrt{2} d_1}{x_k^2 - [n - \frac{1}{2}]^2 \pi^2} +
\frac{\sqrt{2} d_1}{x_k} O(n^{-2})
\end{eqnarray}
\noindent where $d_1$ is a constant.

As shown in \cite{LiHuChenZhang2009}, we have
\begin{equation}
\label{equ:18} J_{1-H}(x_k) = d_2 x_k^{-1/2} + O (x_k^{-3/2})
\end{equation}
\noindent where $d_2$ is a postiche constant.

Thus, based on Eq.(\ref{equ:14}) and (\ref{equ:18}), we have
\begin{eqnarray}
\label{equ:19} & & \sum_{k=1}^{\infty} E \left[z_k^2\right] \hat{\mu}_{n,k}^2 \nonumber \\
& = & \sum_{k=1}^{\infty} \frac{16 c_H^2 d_1^2}{x_k^{2H} J_{1-H}^2 (x_k) \left[ x_k^2 - [n - \frac{1}{2}]^2 \pi^2 \right]^2} + O(n^{-3}) \nonumber \\
& = & d_3 \sum_{k=1}^{\infty} \frac{x_k^{-2H+1}}{\left[ x_k^2 - [n -
\frac{1}{2}]^2 \pi^2 \right]^2} + O(n^{-3}) \nonumber \\
\end{eqnarray}
\noindent where $d_3$ is a positive constant.

It is easy to show that
\begin{eqnarray}
\label{equ:20} \sum_{k=1}^{\infty} \frac{x_k^{-2H+1}}{\left[ x_k^2 -
[n - \frac{1}{2}]^2 \pi^2 \right]^2} > \left.
\frac{x_k^{-2H+1}}{\left[ x_k^2 - [n - \frac{1}{2}]^2 \pi^2
\right]^2} \right \vert_{k=n} = d_3 n^{-2H-1} + O(n^{-3}) \nonumber \\
\end{eqnarray}
and
\begin{eqnarray}
\label{equ:21} \sum_{k=1}^{\infty} \frac{x_k^{-2H+1}}{\left[ x_k^2 -
[n - \frac{1}{2}]^2 \pi^2 \right]^2} < d_3 \sum_{k=1}^{\infty}
\frac{x_k^{-2H+1}}{\left[(2k-1) n \right]^2} + O(n^{-3}) = d_4
n^{-2H-1} + O(n^{-3}) \nonumber \\
\end{eqnarray}
\noindent where $d_3$ and $d_4$ are positive constants.

Noticing that $\sum_{k=1}^{\infty} \frac{x_k^{-2H+1}}{\left[ x_k^2 -
[n - \frac{1}{2}]^2 \pi^2 \right]^2}$ converges, based on
(\ref{equ:13})-(\ref{equ:15}), we have
\begin{equation}
\label{equ:22} \sum_{k=1}^{\infty} E \left[z_k^2\right]
\hat{\mu}_{n,k}^2 = d_5 n^{-2H-1} + O(n^{-3}) \sim n^{-2H-1}
\end{equation}
\noindent where $d_5$ is a positive constant.

Similarly, we can prove that the projection coefficients of $\{
\frac{1 - \cos(y_n t)}{y_n} \}_{n=1}^\infty$ satisfies
\begin{equation}
\label{equ:23} \sum_{k=1}^{\infty} E \left[w_k^2\right]
\tilde{\mu}_{n,k}^2 \sim n^{-2H-1}
\end{equation}
\noindent where $\tilde{\mu}_{n,k} = \int_0^1 \frac{1 - \cos(y_k
t)}{y_k} \sqrt{2} \sin([n - \frac{1}{2}] \pi t) dt$.

Due to the independence of $z_k$ and $w_k$, we have
\begin{equation}
\label{equ:24} \lambda_n = E \left[c_n^2 \right] =
\sum_{k=1}^{\infty} \left( E \left[z_k^2\right] \hat{\mu}_{n,k}^2 +
E \left[w_k^2\right] \tilde{\mu}_{n,k}^2 \right) \sim n^{-2H-1}
\end{equation}

Therefore, our proof in \cite{LiHuChenZhang2009} is repaired.

In summary, the nice proof given by Bronski in
\cite{Bronski2003a}-\cite{Bronski2003b} is a direct attack on the
problem, and our proof detours. However, the appendix in
\cite{LiHuChenZhang2009} plus this response gives another view on
the asympotics of K-L expansion of fBm process and meanwhile shows
how the important results obtained in
\cite{Bronski2003a}-\cite{Bronski2003b} and \cite{DzhaparidzeaZanten2004}-
\cite{DzhaparidzeaZanten2005b} can be linked together.

\section{Some Discussions}
\label{sec:4}

Finally, we would like to discuss the interesting question raised in
\cite{Zanten2010}: \textit{If a centered Gaussian process $G_t$
admits two different series expansions, under what conditions do the
two eigenvalue sequences have the same asymptotic behavior?}

We think this question can be partly solved by evaluating the
mapping operator $T$ between the two sets of basis functions. If the
mapping $T$ consists of \textbf{\textit{appropriate}} projection
coefficients, the asymptotics of the eigenvalues can still be held.

The general cases are obviously too difficult to solve in this short
response. In the follows, we will briefly discuss a special case:
when one basis is a Riesz basis and the other is a orthonormal
basis.

Suppose we have a expansion of the integral kernel $K$ is in
$L_2([0, 1]\times[0, 1])$ on a Riesz basis $\{ \psi_n(t)
\}_{n=1}^\infty$ (no need to be orthogonal) as $K(s, t) =
\sum_{n=1}^\infty \tau \psi_n(s) \psi_n(t)$; and meanwhile we have
the K-L expansion of $K$ on a orthonormal basis $\{ \phi_n(t)
\}_{n=1}^\infty$ in the same space as $K(s, t) = \sum_{n=1}^\infty
\lambda \phi_n(s) \phi_n(t)$.

Based on the property of Riesz basis \cite{Young1980}-
\cite{Christensen2008}, we can always find a linear bounded
bijective operator $T$ satisfying $\{ \psi_n(t) \}_{n=1}^\infty = \{
T \phi_n(t) \}_{n=1}^\infty$.

The basis function used in
\cite{DzhaparidzeaZanten2004}-\cite{DzhaparidzeaZanten2005b} can be
viewed as a special Riesz basis, which satisfying the above
requirement. Thus, we can study the asymptotics of K-L eigenvalues
by using orthogonal projection.

Since $\{ \psi_n(t) \}_{n=1}^\infty$ is a Riesz basis, it will
associate with a set of Riesz sequence. Hence, if $A$ is the
infinite matrix representing this bounded linear operator $T$, the
sequence $\{A_{k,n}\}_{k=1}^\infty$ formed by the $n$th column of
$A$ is a Bessel sequence in $l^2$.

Assume that $\lambda_n$ and $\tau_n$ have the same asymptotics.
According to \cite{Weyl1912}-\cite{FerreiraMenegatto2009}, the upper
bound for the decaying rate of the eigenvalues for a smooth Mercer
kernel is $O(n^{-1})$. Thus, we have $\lambda_n, \tau_n \sim
n^{-p}$, $p > 1$.

Since $\{A_{k,n}\}_{k=1}^\infty$ is a Bessel sequence,
$\sum_{k=1}^\infty A_{k,n}^2 \tau_k \le C_1 \sum_{k=1}^\infty \tau_k
= C_2$ when $\tau_n \sim n^{-p}$, $C_1$ and $C_2$ are constants.
Thus, $\sum_{k=1}^\infty A_{k,n}^2 \tau_k$ converges. Based on the
mapping relation, we have
\begin{equation}
\label{equ:25} \lambda_n = \sum_{k=1}^\infty A_{k,n}^2 \tau_k
\end{equation}
\noindent or equivalently
\begin{equation}
\label{equ:26} d_6 n^{-p} = \sum_{k=1}^\infty A_{k,n}^2 k^{-p} +
O(n^{-p})
\end{equation}
\noindent which indicates that given a $n \in \mathbb{N}$, the
maximum value of $A_{k,n}$ in terms of $k$ locates at a point
$k^{*}$ that is approximately proportional to $n$ (say, $k^{*} =
\lfloor d_7 n \rfloor$). Here $d_6$ and $d_7$ are two positive
constants.

Similarly, if $\{ \psi_n(t) \}_{n=1}^\infty$ and $\{ \phi_n(t)
\}_{n=1}^\infty$ are two different Riesz bases, we can always find
two linear bounded bijective operators $U$ and $V$ satisfying $\{ U
\psi_n(t) \}_{n=1}^\infty = \{ V \phi_n(t) \}_{n=1}^\infty$. Thus,
if $G_t$ admits two series expansions on different Riesz bases, we
may possibly study the asymptotic behavior on one basis from the
knowledge on the asymptotic behaviors of another.

Besides, when $H \rightarrow 0$, the decaying rate of the
eigenvalues will approach the bound $\lambda_n \sim O(n^{-1})$ as
$\lambda_n \sim n^{-2H-1}$. This can be another example in practice
supporting Weyl's conclusion: the rate $O(n^{-1})$ for the
eigenvalues of a smooth Mercer-like kernel cannot be improved in
general \cite{Weyl1912}-\cite{FerreiraMenegatto2009}.


\end{document}